\documentclass[%
aps,
prl,%
amsmath,amssymb,
reprint,
showkeys,
twocolumn]{revtex4-2}

\usepackage{enumitem}
\usepackage{mathrsfs}
\usepackage{makecell}
\usepackage{array}
\usepackage{mathtools}
\usepackage[dvipsnames]{xcolor}
\usepackage{lipsum}
\usepackage{tikz}
\usepackage{wrapfig}
\usepackage[T1]{fontenc}
\usepackage[utf8]{inputenc}
\usepackage{amsmath}
\usepackage{amssymb}
\usepackage{graphicx}
\usepackage{refstyle}
\usepackage{ctable}
\usepackage{multirow}
\usepackage{upgreek}
\usepackage{textgreek}
\usepackage{blkarray,booktabs,bigstrut}

\usepackage[
citecolor=blue,
colorlinks,
linkcolor=blue,
urlcolor=blue,
]{hyperref}

\usepackage{dcolumn}
\usepackage{bm}
\usepackage{tabularx,booktabs}
\usepackage{multirow}
\usepackage{float}
\usepackage{amsmath}
\usepackage{euscript}
\begin{document}
	\title{Stochastic evolutionary dynamics in structured populations: Network degree governs evolutionary stability}
	\author{Vikash Kumar Dubey}%
	\email{vdubey9818@gmail.com}
	\affiliation{%
		Department of Physics,
		Indian Institute of Technology Kanpur,
		Uttar Pradesh 208016, India.
	}%
	\author{Sagar Chakraborty}
	\email{sagarc@iitk.ac.in}
	\affiliation{%
		Department of Physics,
		Indian Institute of Technology Kanpur,
		Uttar Pradesh 208016, India.
	}
	\begin{abstract}
Stochasticity shapes evolutionary dynamics in finite populations beyond the reach of infinite-population theories. Using a stochastic birth–death framework on random regular networks, we establish evolutionary stability criteria for finite structured populations playing three-strategy games and identify critical degree thresholds governing invasion dynamics. Our framework goes beyond two-strategy evolutionary games to reveal that the initial composition of multi-type mutants, together with population structure and stochasticity, shapes the evolutionary fate of the resident population.
	\end{abstract}
	\keywords{Structured populations, Stochastic processes, Fixation probability, Evolutionary stability}
	\maketitle

Real interacting systems are rarely well mixed and infinite; the pattern of connections among individuals---encompassing both structure (who interacts with whom) and connectivity (with how many)---is a defining feature of the system~\cite{WattsStrogatz1998,BarabasiAlbert1999}, shaping outcomes from epidemic spread~\cite{PastorSatorras2001,PastorSatorras2015} to opinion dynamics~\cite{Castellano2009}. In evolutionary games, the underlying finite structure plays a crucial role: Nowak and May~\cite{NowakMay1992} showed that spatial structure alone can rescue cooperation from collapse in the Prisoner's Dilemma (PD) game---a finding later placed on a rigorous footing by evolutionary graph theory~\cite{Lieberman2005,SzaboFath2007}. Moreover, mere connectivity itself can be the decisive parameter: on a regular network of degree $k$, cooperation is favoured when the benefit-to-cost ratio exceeds $k$ ($b/c > k$)~\cite{Ohtsuki2006_Nature,Ohtsuki_2006,TaylorDayWild2007} in PD games.

Within the paradigm of evolutionary game theory~\cite{hofbauer_book,Maynard_Smith1982,Nowak_2006_book,Friedman_Sinervo_Book}, the most important practical question is that of evolutionary stability of a system. This motivated and led to establishment of the concept of ESS~\cite{MaynardSmith1973,Maynard_Smith1982}---evolutionarily stable strategy (or population state)---a cornerstone of evolutionary biology and ecology. Essentially, ESS characterizes the resident population state that is resistant to invasion by mutants. For finite population of size $N$, the concept translates to fulfilling following two criteria by a resident population that is $\textrm{ESS}_N$~\cite{Nowak2004,Nowak_2006_book}: (i) \emph{invasion resistance}---the fitness of resident must be larger than that of the mutant, and (ii) \emph{replacement resistance}---the probability, known as \emph{fixation probability}, that the linage of mutant completely replaces the resident population is less than the fixation probability during random diffusion (known as neutral drift), i.e., when no selection is in action. 

Although the theory of $\textrm{ESS}_N$ has been quite successful, it does not address two important aspects of real evolutionary systems: (i) appearance of more than one type of mutant, and (ii) effect of specific network topology of the population. This paper fills these lacunae by rigorously calculating the analytical expressions of fixation probabilities in the context of three types of individuals on random regular networks, and thence furnishing criteria for evolutionary stability in such hitherto overlooked scenarios.

The curiosity about more than one kind of mutants is not merely academic. Note that in the language of games, $\textrm{ESS}_N$ has been defined only for $2\times 2$ games. However, one is aware of omnipresence of higher dimensional games all around us. Specifically, several $3\times 3$ games, including the \emph{Rock-Scissors-Paper} (RSP), \emph{coordination}, and \emph{Hawk-Dove-Bourgeois} (HDB) games, provide paradigmatic settings for studying such multi-strategy interactions across evolutionary biology and social systems.

The RSP game is a canonical model of cyclic dominance, with applications ranging from biodiversity in natural systems~\cite{Kerr2002,Czrn2002,Kirkup2004,JacksonBuss1975,Sinervo1996} to cyclic competition in social systems~\cite{Frey2013,Szolnoki2014,Wang2014,Demirel2011,Kleshnina2021}.  Mean-field dynamics predict stable coexistence or cyclic oscillations~\cite{Szolnoki2014,Friedman_Sinervo_Book}. Coordination and HDB games provide contrasting examples: Coordination games exhibit multi-stability, with deterministic replicator dynamics~\cite{hofbauer_book,Taylor1978} leading to different stable states depending on initial conditions~\cite{Lewis2013,Young2015,Jackson2014,Keser2012}. The HDB game~\cite{MaynardSmith_Parker_1976,Maynard_Smith1982}, in contrast, introduces role-dependent behaviour through the Bourgeois strategy, which plays Hawk when owning a resource and Dove when intruding.

\begin{figure*}[ht]
	\includegraphics[scale=0.60]{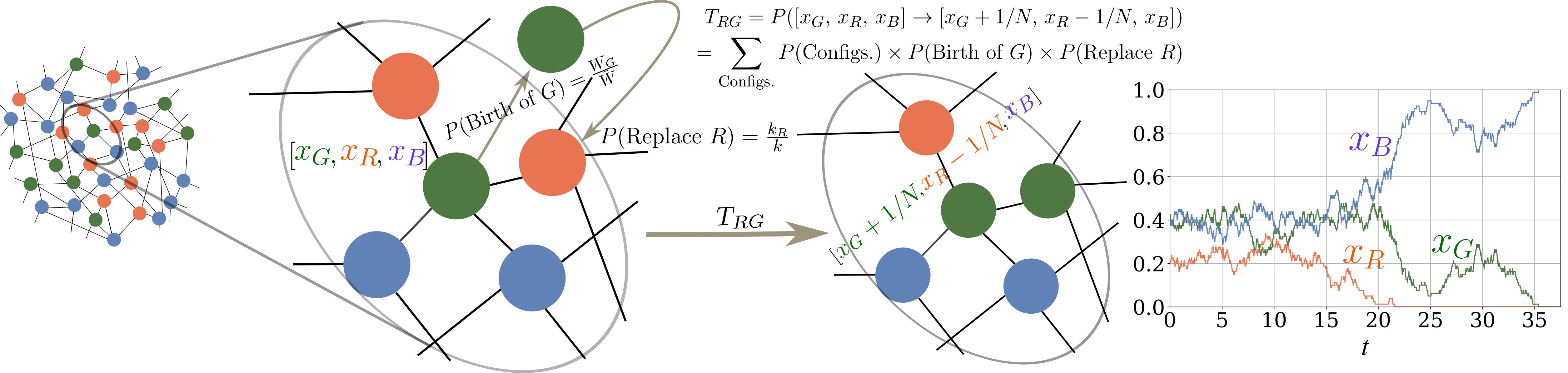}
	\caption{Schematic of birth--death (BD) process on a random regular network with degree $k$: Figure shows a random regular network with $k=4$, containing three types of individuals represented by red (R), blue (B), and green (G) nodes. For illustration, a green individual reproduces with probability proportional to its fitness relative to the mean fitness, i.e., $W_G/\bar W$, and randomly replaces one of its $k$ neighbours (here red neighbuor). The transition probability ($T_{RG}$) of such event of green replacing red depends on three components written in the figure. The rightmost panel illustrates the stochastic changes in the frequencies---$x_R$, $x_B$, and $x_G$---of the three types over time due to repeated occurrence of such BD events.
	}
	\label{fig:schematic}
\end{figure*}

The central challenge in developing the ESS framework in structured finite population is the derivation of fixation probabilities for general three-strategy cases. We overcome this challenge by developing a stochastic framework for a birth--death (BD) process on a random regular network in which population size ($N$) and network connectivity ($k$) serve as tunable parameters in determining the exact nature of the underlying stochastic differential equation (SDE), the resulting fixation probabilities, and the defining criteria of ESS. Although we develop the framework for a BD process on a random regular network, the methodology can be extended to other evolutionary processes.

We begin by considering a random regular network with degree $k$ (see Fig.~\ref{fig:schematic}), consisting of $N$ individuals belonging to three types, represented by red, green, and blue nodes. Let $n_i$ denotes the number of individuals of type~$i$, where $i\in\{1,2,3\}$, so that total number of individuals $N = \sum_i n_i(t)$. In the BD process, a randomly chosen individual reproduces another individual of the same type based on its fitness, and the resulting offspring replaces a randomly chosen neighbour. Since the BD process keeps $N$ fixed, the state of the system can be represented by the frequency vector $\bm{x}=(x_1,x_2,x_3)\equiv (n_1/N,n_2/N,n_3/N)$. Using the constraint $x_1+x_2+x_3=1$, we can express $x_3=1-x_1-x_2$, so the state can be equivalently represented by $\bm{x}=(x_1,x_2)$. The fitness of the $i$th type individual is given in terms of its local neighbourhood arrangements, $[k_1,k_2,k_3]$ ($k_j$ is the number of neighbours of type $j$) and the \emph{payoff matrix} $(a_{ij})$ via the following relation $W_i(k_1,k_2,k_3)=1-\omega+\omega\sum_{l}k_la_{il}$ ($\omega$ is the selection strength).

The \emph{transition probability}, from one state to another, depends on two components: (i) \emph{probability of giving birth}, given by the ratio of the fitness of that individual and the average fitness ($W_i/\bar{W}$) and (ii) \emph{probability of replacement}, given by ratio of number of neighbours of that type to the total number of neighbours ($k_j/k$). Both quantities depend on the local configuration of the network. Therefore, the transition probability is obtained by averaging over all possible local configurations, weighted by their respective probabilities (Fig.~\ref{fig:schematic}). To determine these configuration probabilities in a network, we need to characterize the local neighbourhood of an $i-j$ pair (individual of type $i$ is adjacent to type $j$). In particular, we need $q_{l|ij}$, which denotes the probability that an $l$-type individual is adjacent to the $i$-type individual in such an $i-j$ pair. In general, accounting for these conditional probabilities can be cumbersome. To obtain a closed-form expression, we therefore use the \emph{pair approximation method}~\cite{vanBaalen2000,Matsuda1992,Ohtsuki_2006}, which assumes that the local probabilities depend only on the adjacent player $q_{l|ij}\approx q_{l|i}$.

Using these transition probabilities, one can derive the Fokker--Planck (FP) equation~\cite{Gardiner2009,VanKampen2007} for the probability density $\rho(\bm{x}, t)$ of the system being in state $\bm{x}$ at time $t$ (see Supplementary Material). The details of the BD process and the network connectivity are encoded in the drift and diffusion terms $A_i(\bm{x})$ and $B_{ij}(\bm{x})$, which can be expressed explicitly in terms of $\bm{x}$, $k$, $N$ and the payoff matrix ($a_{ij}$). The closed form expression of the drift term up to $\mathcal{O}(\omega)$ is given as
$A_i(\bm{x})\approx\frac{\omega(k-2)^2}{k-1}x_i(f_i+g_i-\phi)$,
where $f_i=\sum_lx_la_{il}$, $\phi=\sum_{ij}x_ix_ja_{ij}$, $g_i=\sum_lx_jb_{il}$, and $b_{ij}=\frac{a_{ij}+a_{ii}-a_{jj}-a_{ji}}{k-2}$.
The diffusion matrix is written as perturbative expansion in $\omega$: $B\approx B^{(0)}+\omega B^{(1)}$, where the expressions of the perturbation coefficients are given in Appendix. Using this FP equation one can easily find the its corresponding Itô SDE: $d\bm{X} = \mathbf{A}(\bm{X})\, dt + D(\bm{X})\, d\bm{W}_t$, where $D(\bm{X})$ is a $2\times 2$ matrix  given by $\sum_{l=1}^{2}D_{il}D_{lj}=B_{ij}$, and $\bm{W}_t$ is a $2$-dimensional Wiener process~\cite{Gardiner2009,VanKampen2007}.

This is a crucial step, because now we can use the standard approach of solving the backward FP equation to determine that the probability that system fixates at one of the absorbing states~\cite{Ewens2004}. The final expressions for the fixation probabilities, $\upphi_1$ for fixation at the state $(1,0)$ and $\upphi_2$ for fixation at the state $(0,1)$, can be written compactly as functions of initial frequency $(x_1,x_2)$ as,
\begin{eqnarray}\label{eqn:fixation_phi_3by3}
	\upphi_i=x_i+\omega\frac{N(k-2)}{6}\Big[\beta^{(i)}_1x_i+\beta^{(i)}_2x_i^2+\beta^{(i)}_3x_i^3\nonumber\\
	+\beta^{(i)}_4 x_ix_{3-i}+\beta^{(i)}_5x_i^2x_{3-i}+\beta^{(i)}_6 x_ix_{3-i}^2\Big].
\end{eqnarray}
Here $\beta^{(i)}_1=2a_{i3}+3b_{i3}-2a_{33}+a_{ii}-a_{3i}$, $\beta^{(i)}_2=-3a_{i3}-3b_{i3}+3a_{33}$, $\beta^{(i)}_3=-a_{ii}+a_{i3}+a_{3i}-a_{33}$, $\beta^{(i)}_4=2a_{i,3-i}-2a_{i3}+3b_{i,3-i}-3b_{i3}-a_{3-i,3}-a_{3,3-i}+3a_{33}+a_{3i}-a_{3-i,i}-a_{3-i,3-i}$, $\beta^{(i)}_5=-a_{i,3-i}+a_{i3}-a_{3-i,i}+a_{3-i,3}+a_{3i}+a_{3,3-i}-2a_{33}$ and $\beta^{(i)}_6=-a_{3-i,3-i}+a_{3-i,3}+a_{3,3-i}-a_{33}$, where $i\in\{1,2\}$ (see Supplementary Material).  The third fixation towards $(0,0)$ state is simply $\upphi_3=1-\sum_{i=1}^2\upphi_i$.

Since the fixation probability given in Eq.~(\ref{eqn:fixation_phi_3by3}) holds for arbitrary frequencies of the three types, we can ask questions that lie beyond the scope of traditional $\mathrm{ESS}_N$ theory~\cite{Nowak_2006_book}, defined for two types, with one type conventionally taken to be a single rare mutant. In the current setting, we have greater flexibility: one type can be treated as the dominant resident whose evolutionary stability is of interest, while the remaining two types constitute a total  $\mu$ number of mutants, distributed between them in the ratio $\gamma:(1-\gamma)$ (see also Ref.~\cite{Boyd1987,Boyd1989,Gerrish1998}). The single-mutant case is recovered as the special cases by putting $\gamma=0$ or $1$.

Strictly speaking, two distinct mutant types need not arise at exactly the same instant for this framework to be relevant: in a more detailed model, a second mutant lineage typically arises while the first is still segregating at low frequency, rather than the two appearing simultaneously~\cite{Gerrish1998,Desai2007,Yu2010}. Since this formalism, like other conventional formulations, does not track the time of appearance of the mutants, we can pose a well-defined snapshot question: at a moment when both rare types are simultaneously present, is the resident protected against invasion? This is a very realistic simplification as simultaneous presence of multiple rare/deviant types alongside a resident type is well documented empirically and theoretically, both in biological~\cite{Boyd1987,Boyd1989,Gerrish1998,Gerrish1998,Desai2007,Yu2010,Drake2005,Drake2007,Marusyk2014,Beckman2017,Acar2020,Makau2022,DiazCaballero2023} and in social systems~\cite{Holley1975,Vazquez2004,Xie2011,Bhme2012,Waagen2015,HerrerasAzcu2019,Redner2019}.

With the setup for two types of mutants in place, we can now ask how selection shapes their evolutionary fate. It is important to note that, in finite populations, mutants can reach fixation even in the absence of selection, purely through neutral drift~\cite{Kimura1957,Kimura1962}. Under neutral drift, the fixation probabilities of the two mutant types are determined by their respective initial abundances, i.e., $\gamma\mu/N$ and $(1-\gamma)\mu/N$. This provides a neutral baseline for evolutionary dynamics. The natural question, then, is how selection combines with structure to modify this baseline: does it enhance or suppress the probability of mutant fixation relative to neutral drift?

The weak-selection fixation probabilities in Eq.~(\ref{eqn:fixation_phi_3by3}) quantify how network structure modifies this neutral fixation baseline. Without loss of generality, let type 3 be the resident and types 1 and 2 be the mutants. To leading order, their fixation probabilities are given by $\upphi_1=\gamma\mu/N+\mathcal{O}(\omega)$ and $\upphi_2=(1-\gamma)\mu/N+\mathcal{O}(\omega)$, respectively using Eq.~(\ref{eqn:fixation_phi_3by3}). The corrections of order $\omega$ capture the effect of selection, with network degree governing whether it promotes or suppresses mutant fixation. Therefore, replacement resistance requires the fixation probability of each mutant type to remain below its neutral expectation, ensuring that the resident remains protected against replacement in this probabilistic sense:
$\upphi_1(\mu\gamma/N,\mu(1-\gamma)/N)<\mu\gamma/N$ and $\upphi_2(\mu\gamma/N,\mu(1-\gamma)/N)<\mu(1-\gamma)/N$.

As discussed earlier, an $\mathrm{ESS}_{N,k}$ must also satisfy the invasion resistance criterion, which requires the resident strategy to be fitter than each of the invading mutant. The expected fitness of type $i$, $F_i$, in the network is obtained by averaging $W_i$ over $\sum_{k_1+k_2+k_3=1} k!/(k_1!k_2!k_3!)$ possible configurations of the neighbourhoods: $F_i = 1-\omega+\omega\frac{k(k-2)}{k-1}\sum_l x_l a_{il} + \frac{k}{k-1}\omega a_{ii}$ (see Supplementary Material). Thus, for type 3 to be invasion resistant, we require $F_3>F_1$ and $F_3>F_2$. 
Together, these two conditions ensure that the resident is protected against invasion by either mutants. In summary, the definition of ESS in this context naturally follows:

\emph{In a structured population of size $N$ and degree $k$, suppose that $\mu$ mutants are present, where type 1 and type 2 are in the ratio $\gamma:(1-\gamma)$, type 3 is an $\mathrm{ESS}_{N,k}$ if the
following two conditions are satisfied, $\forall i\in\{1,2\}$:
\begin{itemize}
	\item[(i)] Invasion resistance:
	$N\big[(a_{33}-a_{ii})+(k-2)(a_{33}-a_{i3})\big]+ (k-2)\mu\big[ \gamma(a_{31}-a_{i1}-a_{32}+a_{i2}) +(a_{32}-a_{i2}-a_{33}+a_{i3})\big]>0$,
	\item[(ii)] Replacement resistance: $N^2(k\EuScript{A}^{(i)}_4-\EuScript{A}^{(i)}_1)+N\mu(k\EuScript{A}^{(i)}_5(\gamma)-\EuScript{A}^{(i)}_2(\gamma)
	)+(k-2)\mu^2\EuScript{A}^{(i)}_3(\gamma)<0$;
\end{itemize}
where, $\EuScript{A}^{(i)}_1=(a_{i3}-a_{33})+(a_{3i}-a_{ii})$,~ $\EuScript{A}^{(i)}_2(\gamma)=\gamma^{2-i}(1-\gamma)^{i-1}(-3a_{i3}+3a_{33}+3a_{ii}-3a_{3i})+\gamma^{i-1}(1-\gamma)^{2-i}(a_{3-i,i}+a_{i,3-i}+a_{3-i,3-i}-a_{i3}-a_{3i}+3a_{33}-2a_{3-i,3}-2a_{3,3-i})$,~
$\EuScript{A}^{(i)}_3(\gamma)=(\gamma^{2-i}(1-\gamma)^{i-1})^2\beta^{(i)}_3+\gamma(1-\gamma)\beta^{(i)}_5+(\gamma^{i-1}(1-\gamma)^{2-i})^2\beta^{(i)}_6$,~
$\EuScript{A}^{(i)}_4=2(a_{i3}-a_{33})+a_{ii}-a_{3i}$,~and~ $\EuScript{A}^{(i)}_5(\gamma)=\gamma^{2-i}(1-\gamma)^{i-1}(-3a_{i3}+3a_{33})+\gamma^{i-1}(1-\gamma)^{2-i}(2a_{i,3-i}-2a_{i3}-a_{3-i,3}-a_{3,3-i}+3a_{33}+a_{3i}-a_{3-i,i}-a_{3-i,3-i})$}. The calculational details of above mathematical expressions can be found in Supplementary Material. For analogous results in 2-type systems, refer Appendix.

The definition of $\mathrm{ESS}_{N,k}$ makes explicit that network connectivity is a prominent parameter that can alter the evolutionary fate of a resident strategy. In particular, we identify four critical degrees, $k^{(1)}_\mathrm{I}$, $k^{(2)}_\mathrm{I}$, $k^{(1)}_\mathrm{R}$, and $k^{(2)}_\mathrm{R}$, corresponding respectively to the two invasion resistance criteria and the two replacement resistance criteria (the expressions are given in the Appendix). These critical degrees mark the boundaries between the regions of the $(N,k)$ space in which the resident is, and is not, favoured. As the network degree $k$ varies, the point in the $(N,k)$ space moves relative to these boundaries, so that crossing one of them---call it the effective critical degree---switches the resident between being an ESS$_{N,k}$ and not being one, and vice versa. Which critical degree is effective, and the direction of this transition (from ESS$_{N,k}$ to not being one, or from not being an ESS$_{N,k}$ to being one), depend on the underlying game.

\begin{figure}[ht]
	\includegraphics[scale=0.98]{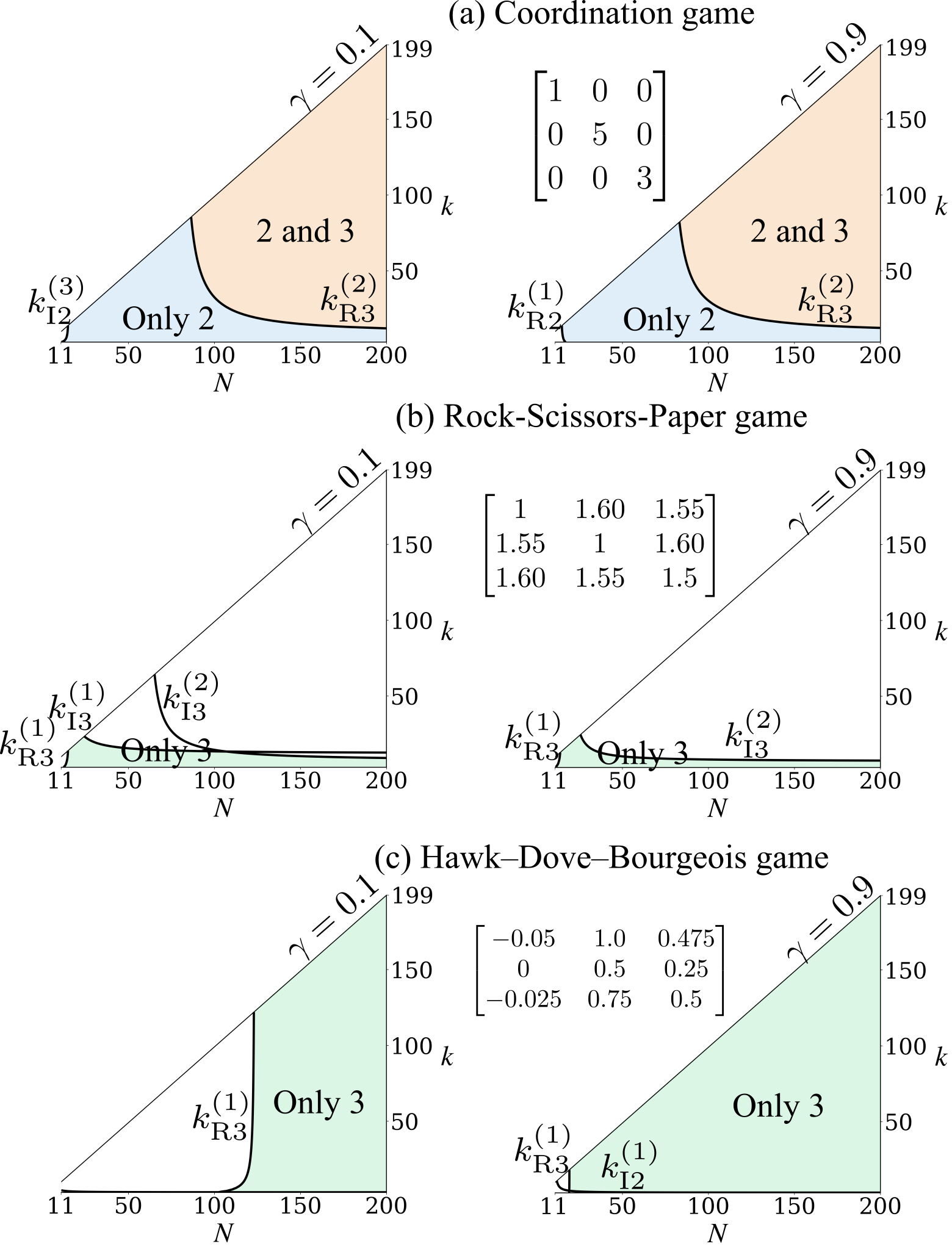}
	\caption{Pictorial demonstration of $\mathrm{ESS}_{N,k}$ and critical degrees for the coordination, the RSP, and the HDB games in the $(N,k)$ space. The white region represents the absence of any $\mathrm{ESS}_{N,k}$, while the light blue, light green, and light-orange regions represent parameter regimes in which only strategy 2, only strategy 3, and both strategies 2 and 3, respectively, are $\mathrm{ESS}_{N,k}$. The figure shows only those $k^{(l)}_{\mathrm{I}j}$ and $k^{(l)}_{\mathrm{R}j}$ that form the boundaries of the regions satisfying the $\mathrm{ESS}_{N,k}$ conditions (i.e., the effective critical degrees); the other critical degrees are not shown for clarity. See the Supplementary Material for all critical degrees.}
	\label{fig:3by3_game}
\end{figure}

Having developed this general framework, we now apply it to the some well-known games discussed earlier, namely the coordination, the RSP, and the HDB games to investigate whether the evolutionary stabilities in unstructured populations are modified due to network-structure. To analyze this systematically, we consider one strategy as the resident and the other two as mutants. While we vary $N$ from 11 to 200, and hence, $k$ from 3 to $N-1$; for illustrative purpose, the initial total mutant number  and the initial mutant ratio are, respectively, chosen to be $\mu=10$ and $\gamma=0.1\,\&\,0.9$. Now, to completely analyze the system we check the ESS$_{N,k}$ condition for all the three strategies in each game. We use the subscript $j\in\{1,2,3\}$ in $k^{(l)}_{Ij}$ and $k^{(l)}_{Rj}$ to denote the strategy which is taken as resident. Fig.~\ref{fig:3by3_game} presents the resulting $\mathrm{ESS}_{N,k}$ regions for the coordination, the RPS, and the HDB games.

As shown in Fig.~\ref{fig:3by3_game}(a), in the coordination game, the pure strategy 3 is not an $\mathrm{ESS}_{N,k}$ when $k<k^{(2)}_{\rm R3}(N)$. More generally, when $k<k^{(2)}_{\rm R3}(N)$, only the strategy with the highest self-payoff, i.e., the payoff $a_{ii}$ obtained by individual of type $i$ when interacting with another individual of the same type, is an $\mathrm{ESS}_{N,k}$. This has no counterpart in unstructured infinite population ESS theory, where each pure strategy is an ESS and therefore cannot be invaded by rare mutants. In a finite and structured population, however, mutants may collectively invade when the network degree $k$ is sufficiently low. Pure strategies in coordination games are commonly used to model conventions (e.g., a norm or standard of behavior)~\cite{Lewis_1969,Young_1993}, which are stable equilibria under global, well-mixed interaction. Our result suggests that when individuals interact with only a small number of neighbours, a cluster of mutants can collectively invade and overturn an established convention~\cite{Ellison_1993,Ellison_2000}.

A general RSP game is non-transitive, however the non-transitivity leave open a choice, whether the resident strategy or the remaining non-dominant strategy is the second-best response (later case is called an apostatic RSP game~\cite{Friedman_Sinervo_Book}). Furthermore, real biological RSP systems generically have unequal diagonal payoffs~\cite{Sinervo1996,Friedman_Sinervo_Book,Neff2003,Yeates2016}. Motivated by this, while keeping non-tranisitivity, we break the diagonal symmetry of the apostatic RSP payoff matrix, assigning strategy 3 a higher self-payoff as compare to the other two strategies, i.e., $a_{33}>a_{11},a_{22}$ (Fig.~\ref{fig:3by3_game}(b)). Despite the absence of a pure ESS in the unstructured infinite population theory, strategy 3 can become an $\mathrm{ESS}_{N,k}$ in a finite and structured population when $k$ is below the joint boundary determined by $k^{(1)}_{\rm I3}$ and $k^{(2)}_{\rm I3}$, as shown in Fig.~\ref{fig:3by3_game}(b). Thus, ``paper'' can resist invasion by ``rock'' and ``scissors'' collectively when the network connectivity is sufficiently low. This phenomenon also has no counterpart in unstructured infinite population models. Our results therefore suggest a possible mechanism by which a single type can attain dominance in finite structured populations in RSP games, consistent with observations that single-type dominance often occurs in natural systems~\cite{Park2020,Liao2020}.

In the HDB game, when the cost of escalation exceeds the benefit, the Bourgeois strategy is an ESS and Hawk cannot invade~\cite{Maynard_Smith1982} under the standard well-mixed infinite population theory. However, as shown in Fig.~\ref{fig:3by3_game}(c), this result also depends on the location in the $(N,k)$ parameter space. For $k<k^{(1)}_{\rm R3}$, the Bourgeois strategy is no longer an $\mathrm{ESS}_{N,k}$, indicating that conditional strategy like Bourgeois can be invaded when network connectivity falls below the corresponding cutoff. This result has a natural interpretation in territorial species, where individuals contest for resources with only a limited set of spatial or social neighbours~\cite{MaynardSmith_Parker_1976}. When this neighbourhood is small, the Hawk mutants can avoid costly escalated fights and dove mutants can avoid being repeatedly displaced by resident (Bourgeois). On the other hand, in the large, well-mixed limit, each individual's payoff is averaged over interactions with a large number of residents, diluting this local advantage. This suggests that the stability of ownership-based contest resolution is itself sensitive to the spatial or social structure of the population, not just to the cost-benefit ratio of escalation.

Taken together, these three examples demonstrate that the ability of a strategy to resist invasion by mutants depends systematically on network connectivity, a dependence that is rigorously captured by the $\mathrm{ESS}_{N,k}$ framework developed here.

 Mutations favoured by selection can still be lost to chance, and disfavoured ones can still spread to fixation—in finite populations, stochastic fluctuations are not a minor correction to selection but a force that can override it entirely~\cite{Kimura1957,Kimura1962,Moran1958}. Structure---determined by spatial proximity~\cite{Nowak1992,Szab2007,Nowak2010}, social ties~\cite{Dbarre2014,Kurvers2014}, or contact networks~\cite{Keeling2005,Meyers2006,Newman2002}---adds another layer of complexity and can significantly alter the evolutionary fate of mutants. These effects have no counterpart in infinite-population, deterministic descriptions, and studying the interplay between stochasticity and population structure from an evolutionary perspective is a central objective of this work.

In the course of our study, we have developed a full stochastic framework for BD processes in structured populations, deriving the corresponding FP equation, SDE, and fixation probabilities, allowing for developing the concept of evolutionary stability in structured populations.  More importantly, the examples in Fig.~\ref{fig:3by3_game} reveal that network connectivity is not merely a quantitative modifier of evolutionary outcomes: As the network degree crosses a critical value, it can drive a qualitative transition between evolutionarily favourable and unfavourable configurations even when the underlying payoff structure and population size remain unchanged. Connectivity therefore emerges as a second axis, alongside population size, along which the direction of selection can change.

It is worth emphasizing that by providing analytical expressions of SDEs and fixation probabilities, the framework developed here offers a substantial computational advantage because, as discussed in Ref.~\cite{Traulsen2012}, alternatively using agent-based simulations requires significantly more computational time. 

Our framework potentially opens many unexplored avenues of research in a broad class of finite structured populations. An immediate direction is to extend this approach to other evolutionary processes, such as death--birth and imitation. Incorporating migration, and mutation during reproduction~\cite{Traulsen2012} would provide further avenues for investigation. More broadly, we view the framework developed here as a foundation for systematically exploring stochastic evolutionary dynamics in finite structured populations across a wide range of population structures, update processes, and evolutionary scenarios.

\subsection*{Supplementary Information}
Supplementary Information will be made available upon publication or upon request to the first author.

\bibliography{Dubey_etal_bibliography.bib}
\onecolumngrid
\appendix
\section{Diffusion matrix and critical degrees mathematical expressions}\label{sec:cefficients}
As discussed in the main text, the diffusion matrix can be written as $B\approx B^{(0)}+\omega B^{(1)}$, whose zeroth- and first-order terms are given below (see Supplementary Material for the derivation):
\begin{eqnarray}\label{eqn:difussion_2d_1}
B^{(0)}(\bm{x})&=&	\frac{2(k-2)}{N(k-1)}\begin{bmatrix}
		x_1(1-x_1)& -x_1x_2\\
		-x_1x_2&x_2(1-x_2)
	\end{bmatrix}\\
B^{(1)}_{ii}(\bm{x})&=&\frac{(k-2)^2}{N(k-1)^2} \big[kx_i\left(f_i+\psi_i-\phi\right)
	+(3k-2)x_i^2(\phi-f_i-\xi_i)\big],~i\in\{1,2\}\\
	B^{(1)}_{12}(\bm{x})&=&-\frac{(k-2)}{2N(k-1)^2}\Big[\{(k-2)\phi+\chi\}\Omega_k(\bm{x})+(k-1)G[{\bm f}]
	+G[\text{diag}(a)]-G[{\bm h}]\nonumber\\
	&&+2(k-1)x_1x_2\left\{(k-2)(f_1+f_2)+\sigma_a\right\}\Big]\label{eqn:difussion_2d_2}.~~~~
\end{eqnarray}
Here, $\psi_{i}=\sum_lx_lc_{il}$, $h_i=\sum_lx_la_{li}$,~$\xi_i=\sum_lx_ld_{il}$ ,~$c_{il}=\frac{-(k-2)a_{il}+(k-2)a_{ii}-ka_{ll}+ka_{li}}{k(k-2)}$, $d_{il}=\frac{-(k-2)a_{il}+(5k-6)a_{ii}-(3k-2)a_{ll}-(k-2)a_{li}}{(k-2)(3k-2)}$,~$\chi=\sum_lx_la_{ll}$, $\Omega_k(\bm{x})=(k-1)(2x_1+2x_2-1)-2kx_1x_2$,~$G[{\bm z}]= (-x_1 z_1 - x_2 z_2 + x_3 z_3) + (k-2)(-x_1^2 z_1 - x_2^2 z_2 + x_3^2 z_3)$,  ${\bm f}=(f_1,f_2,f_3)$, ${\bm h}=(h_1,h_2,h_3)$,~$\text{diag}(a)=(a_{11},a_{22}, a_{33})$, and $\sigma_a = a_{11}+a_{22}+a_{12}+a_{21}$.

The expressions of critical degree corresponding to invasion and replacement resistance criteria are the following:
\begin{eqnarray}\label{eqn:critical_kcI_3by3}
	k^{(i)}_\mathrm{I}=\frac{N(a_{33}-2a_{i3}+a_{ii})+2\mu\EuScript{R}^{(i)}(\gamma)}{
		N(a_{33}-a_{i3})+\mu\EuScript{R}^{(i)}(\gamma)},~~~k^{(i)}_\mathrm{R}=\frac{N^2\EuScript{A}^{(i)}_1
		+N\mu \EuScript{A}^{(i)}_2(\gamma)
		+2\mu^2\EuScript{A}^{(i)}_3(\gamma)}{N^2\EuScript{A}^{(i)}_4
		+N\mu\EuScript{A}^{(i)}_5(\gamma)
		+\mu^2\EuScript{A}^{(i)}_3(\gamma)}.
\end{eqnarray}
Here, $\EuScript{R}^{(i)}(\gamma)=\gamma(a_{31}-a_{i1}-a_{32}+a_{i2})+(a_{32}-a_{i2}-a_{33}+a_{i3})$, and the the coefficients $\EuScript{A}^{(i)}_j(\gamma)$ are given in the main text with the definition of ESS$_{N,k}$.

\section{Birth--Death process, ESS$_{N,k}$ and critical degrees for two phenotypes scenarios}\label{sec:appendix_2by2}
For two-strategy games, the expressions simplify because there is only one independent variable. We set $x_1=x$, and hence $x_2=1-x$ and $x_3=0$. Consequently, there is only one component of $A_i(\mathbf{x})$, which we denote by $A(x)$, whose expression is given as $A(x)=\frac{\omega(k-2)^2}{k-1}x(f(x)+g(x)-\phi)$
where $f(x)=xa_{11}+(1-x)a_{12}$, $\phi=\sum_{ij}x_ix_ja_{ij}$, $g(x)=xb_{11}+(1-x)b_{12}$, and $b_{ij}=\frac{a_{ij}+a_{ii}-a_{jj}-a_{ji}}{k-2}$. Similarly, there is only one independent component of the diffusion term $B_{ij}(\mathbf{x})$, which we denote by $B(x)$ and given as:
\begin{eqnarray}\label{eqn:difussion_1d}
	B(x)=\frac{2(k-2)}{N(k-1)}x(1-x)+\omega \frac{k(k-2)^2}{N(k-1)^2}\big[ kx\left(f+\psi-\phi\right)
	+(3k-2)x^2\left(\phi-f-\xi\right)\big],
\end{eqnarray}
where, $\psi=\sum_lx_lc_{1l}$ and $\xi=\sum_lx_ld_{1l}$. These $(c_{ij})$ and $(d_{ij})$ matrices are given in terms of payoff matrix $a_{ij}$ and degree of the regular network $k$, as follows:~$c_{il}=\frac{-(k-2)a_{il}+(k-2)a_{ii}-ka_{ll}+ka_{li}}{k(k-2)}$ and $d_{il}=\frac{-(k-2)a_{il}+(5k-6)a_{ii}-(3k-2)a_{ll}-(k-2)a_{li}}{(k-2)(3k-2)}$. Using these expression we can find the corresponding SDE as the following: $dx = A(x) dt + D(x)\, dW_t$ here $W_t$ is one dimensional Wiener process and $D(x)$ is the square root of $B(x)$.

For two-strategy cases, there are only two absorbing states, namely $x=0$ and $x=1$. Once the stochastic birth--death process reaches an absorbing state, where all individuals are of the same type, the system remains in that state. To obtain the fixation probability at $x=1$, we can either use the drift and diffusion terms derived above and solve the backward Fokker--Planck equation (see Supplementary Material), or directly use Eq.~(\ref{eqn:fixation_phi_3by3}) by setting $(x_1,x_2,x_3)=(x,1-x,0)$. This gives the following expression~(cf.~\cite{Ohtsuki2006_Nature}) for the fixation probability at $x=1$:
\begin{eqnarray}\label{eqn:fixation_prob}
	\Phi(x)=x+\omega\frac{ N x(1-x)}{6}\Big[(k-2)(a_{11}-a_{12}-a_{21}+a_{22})x
	+(k+1)(a_{11}-a_{21})+(2k-1)(a_{12}-a_{22})\Big].
\end{eqnarray}
The fixation probability at $x=0$ is simply given as $1-\Phi(x)$.

Using this fixation expression together with the payoff comparison described in the main text, we can now define $\mathrm{ESS}_{N,k}$ for two-strategy games in finite and structured populations. To determine whether strategy 2 is an $\mathrm{ESS}_{N,k}$, we require both invasion resistance, $F_2>F_1$, and replacement resistance, $\Phi(1/N)<1/N$. Note that, for comparison with the standard literature, we set the number of mutants to $\mu=1$; however, the same arguments can be readily extended to arbitrary $\mu$. \\
\emph{In a structured population of size $N$ and degree $k$, suppose that a mutant of type 1 is present, type 2 is an $\mathrm{ESS}_{N,k}$ if the following two conditions are satisfied:
\begin{itemize}
	\item[](i) Invasion resistance:
	$N\Big[(k-1)a_{22} -a_{11} - (k-2)a_{12}\Big]
	\;+\; (k-2)\Big[(a_{12}+a_{21}) - (a_{11}+a_{22})\Big] \;>\; 0$\\
	(ii) Replacement resistance:
	$N\left[(k+1)(a_{11}-a_{21}) + (2k-1)(a_{12}-a_{22})\right]+(k-2)(a_{11}-a_{12}-a_{21}+a_{22})<0$.
\end{itemize}
}
We can again calculate the two critical degrees using the above ESS$_{N,k}$ condition as follows
\begin{subequations}
\begin{eqnarray}
k_{\rm I}(N)&&=\frac{N(a_{22}+a_{11}-2a_{12}) + 2(a_{12}+a_{21}-a_{11}-a_{22})}{N(a_{22}-a_{12}) + a_{12}+a_{21}-a_{11}-a_{22}},\\	
k_{\rm R}(N) &&=\frac{(N-2)[(a_{21}-a_{11}) + (a_{12}-a_{22})]}{N[(a_{11}-a_{21}) + 2(a_{12}-a_{22})]+(a_{11}-a_{12}-a_{21}+a_{22})}.
\end{eqnarray}
\end{subequations}

\begin{figure}[ht]
	\includegraphics[scale=0.75]{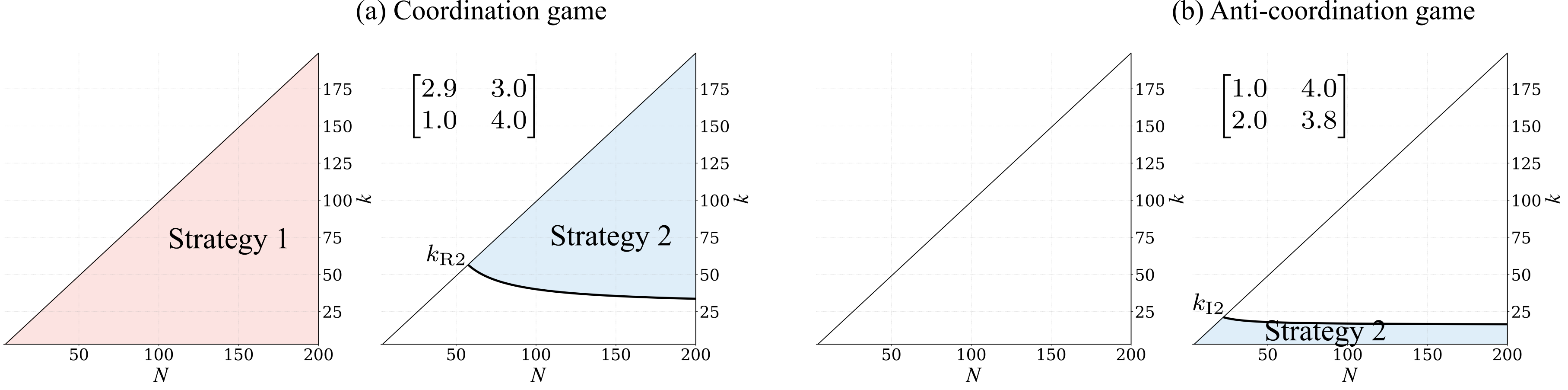}
	\caption{Pictorial demonstration of ESS$_{N,k}$ and critical degrees for coordination and anti-coordination games in $(N,k)$ space: The figure shows the ESS$_{N,k}$ condition for each strategy separately. The light-red and light-blue regions indicate where strategy 1 and strategy 2, respectively, satisfy the ESS$_{N,k}$ condition, while the white region indicates that neither strategy is ESS$_{N,k}$. The black lines denote the critical-degree lines, where the subscript $j$ in $k_{\mathrm{R}j}$ and $k_{\mathrm{I}j}$ indicates the strategy taken as the resident strategy.}
	\label{fig:critical_k_2by2}
\end{figure}

For demonstration of the ESS$_{N,k}$ conditions and effect of critical degree on two-strategy games we plot ESS$_{N,k}$ region in Fig.~\ref{fig:critical_k_2by2} for a coordination and an anti-coordination game. 
As we can see in Fig.~\ref{fig:critical_k_2by2}, in the coordination game, strategy 2 is ESS$_{N,k}$ only when $k>k_{\rm{R}2}$, whereas in the anti-coordination game, pure strategy 2 is ESS$_{N,k}$ when $k<k_{\rm{I}2}$. These transitions cannot be captured by unstructured infinite population theory, where both pure strategies are ESSes in the coordination game, whereas the mixed strategy is the ESS in the anti-coordination game.

\end{document}